%% file: main.tex
\pdfoutput=1

\documentclass[USenglish,cleveref,thm-restate]{lipics-v2021}

\usepackage{centernot}
\usepackage{amssymb}
\usepackage{algorithm2e}
\usepackage{stmaryrd}
\usepackage{bm}
\usepackage{tikz}
\usetikzlibrary{shapes,calc,arrows,automata}
\usepackage{multirow}
\usepackage{array}
\usepackage{mathtools}
\usepackage{proof}
\usepackage{hhline}
\usepackage{stackengine, graphicx}
\usepackage{cite}
\usepackage{microtype}
\usepackage[font=small]{caption}
\usepackage{xspace}
\usepackage{marvosym}

\everypar{\looseness=-1}
\allowdisplaybreaks[4]
\nolinenumbers

\DontPrintSemicolon

\hypersetup{
  pdftitle={Proving Non-Termination by Acceleration Driven Clause Learning},
  pdfauthor={Florian Frohn and Jürgen Giesl},
  colorlinks=true,
  linkcolor=blue,
  citecolor=olive,
  filecolor=magenta,
  urlcolor=cyan
}

\hideLIPIcs  

\DeclareMathAlphabet{\mathpzc}{OT1}{pzc}{m}{it}

\renewcommand{\epsilon}{\varepsilon}
\let\oldphi\phi
\let\oldvarphi\varphi
\renewcommand{\varphi}{\oldphi}
\renewcommand{\phi}{\oldvarphi}

\newcommand{\nt}{\mathsf{nt}}

\newcommand{\tool}[1]{\textsf{#1}}
\newcommand{\inc}[1]{#1\raisebox{.4ex}{\tiny\bf++}}
\newcommand{\dec}[1]{#1\raisebox{.4ex}{\tiny\bf--\! --}}
\newcommand{\eq}[1]{\overset{=}{#1}}

\newcommand{\Init}{\textsc{Init}\xspace}
\newcommand{\Step}{\textsc{Step}\xspace}
\newcommand{\Accelerate}{\textsc{Accelerate}\xspace}
\newcommand{\Covered}{\textsc{Covered}\xspace}
\newcommand{\Backtrack}{\textsc{Backtrack}\xspace}
\newcommand{\Refute}{\textsc{Refute}\xspace}
\newcommand{\Nonterm}{\textsc{Nonterm}\xspace}

\newcommand{\mbp}{\mathsf{sip}}

\newcommand{\unsafe}{\mathsf{unsafe}}

\newcommand{\bt}{\mathsf{bt}}
\newcommand{\cond}{\mathsf{cond}}
\newcommand{\guard}[1]{\,\llbracket #1 \rrbracket}
\newcommand{\accel}{\mathsf{accel}}
\newcommand{\chain}{\mathsf{chain}}
\newcommand{\init}{\mathsf{init}}
\newcommand{\err}{\mathsf{err}}
\newcommand{\QF}{\mathsf{QF}}
\renewcommand{\AA}{\mathcal{A}}
\newcommand{\LL}{\mathcal{L}}
\newcommand{\VV}{\mathcal{V}}
\newcommand{\state}[1]{\ensuremath{\mathfrak{#1}}}

\newcommand{\NN}{\mathbb{N}}
\newcommand{\CC}{\mathcal{C}}

\newcommand{\TT}{\mathcal{T}}

\renewcommand{\SS}{\mathcal{S}}

\newcommand{\Def}{\mathrel{\mathop:}=}

\renewcommand{\emptyset}{\varnothing}

\crefname{equation}{eq.}{equations}%
\crefname{chapter}{chapter}{chapters}%
\crefname{section}{sect.}{sections}%
\crefname{appendix}{app.}{appendices}%
\crefname{enumi}{item}{items}%
\crefname{footnote}{footnote}{footnotes}%
\crefname{figure}{fig.}{figures}%
\crefname{table}{table}{tables}%
\crefname{theorem}{thm.}{theorems}%
\crefname{lemma}{lemma}{lemmas}%
\crefname{corollary}{cor.}{corollaries}%
\crefname{proposition}{proposition}{propositions}%
\crefname{definition}{def.}{definitions}%
\crefname{result}{result}{results}%
\crefname{example}{ex.}{examples}%
\crefname{remark}{remark}{remarks}%
\crefname{note}{note}{notes}%

\title{Proving Non-Termination by Acceleration Driven Clause Learning}
\titlerunning{Proving Non-Termination by Acceleration Driven Clause Learning}
\author{Florian Frohn}{\and \url{https://ffrohn.github.io/}}{florian.frohn@cs.rwth-aachen.de}{https://orcid.org/0000-0003-0902-1994}{}
\author{Jürgen Giesl}{LuFG Informatik 2, RWTH Aachen University, Aachen, Germany \and \url{https://verify.rwth-aachen.de/giesl/}}{giesl@cs.rwth-aachen.de}{https://orcid.org/0000-0003-0283-8520}{}
\authorrunning{F.\ Frohn, J.\ Giesl}

\Copyright{Florian Frohn and Jürgen Giesl}

\relatedversiondetails{See \cite{report}. Full version, including all proofs}{https://arxiv.org/abs/2304.10166} 

\funding{funded by
    the Deutsche Forschungsgemeinschaft (DFG, German Research Foundation)
    - 235950644 (Project GI 274/6-2)}



\begin{document}

\maketitle

\input{abstract}
\input{introduction}
\input{preliminaries}
\input{calculus}
\input{experiments}

\bibliographystyle{plainurl}
\bibliography{refs,crossrefs,strings}

\end{document}

%% file: abstract.tex
\begin{abstract}
  We recently proposed \emph{Acceleration Driven Clause Learning} (ADCL), a
  novel calculus to analyze satisfiability of \emph{Con\-strained Horn Clauses} (CHCs).
  Here, we adapt ADCL to disprove termination of transition systems, and we evaluate its implementation in our tool \tool{LoAT} against the state of the art.
\end{abstract}

%% file: introduction.tex
\section{Introduction}
\label{sect:Introduction}

We are concerned with \emph{dis}proving termination of \emph{transition systems} (TSs), a popular intermediate representation for verification of programs written in more expressive languages.

\begin{example}
  \label{ex:leading}
  Consider the TS $\TT$, where $x',y',z'$ represent the updated values of $x,y,z$, and
  $\eq{x}, \inc{x}, \dec{x}$ abbreviate $x' = x$, $x' = x + 1$, and $x' = x - 1$.
  The first two transitions are a variant of {\tt chc-LIA-Lin\_052} from the \emph{CHC Competition~'22}
    (\url{https://chc-comp.github.io}) and the last two are a variant of {\tt flip2\_rec.jar-obl-8} from \emph{TermComp} \cite{termcomp}.
  \begin{align}
    \init & {} \to \ell_1 \guard{x' \leq 0 \land z' \geq 5000 \land y' \leq z'}  \label{eq:ex1-init} \tag{\protect{\ensuremath{\tau_{\mathsf{i}}}}} \\
    \ell_1 & {} \to \ell_1 \guard{y \leq 2 \cdot z \land \inc{x} \land ((x < z \land \eq{y}) \lor (x \geq z \land \inc{y})) \land \eq{z}} \label{eq:ex1-rec} \tag{\protect{\ensuremath{\tau_{\ell_1}}}} \\
    \ell_1 & {} \to \ell_2 \guard{x = y \land x > 2 \cdot z \land \eq{x} \land \eq{y}} \label{eq:ex1-nonrec} \tag{\protect{\ensuremath{\tau_{\mathsf{\ell_1 \to \ell_2}}}}} \\
    \ell_2 & {} \to \ell_2 \guard{x = y \land x > 0 \land \eq{x} \land \dec{y}} \label{eq:ex1-eq} \tag{\protect{\ensuremath{\tau_{\ell_2}^=}}}\\
    \ell_2 & {} \to \ell_2 \guard{x > 0 \land y > 0 \land x'=y \land ((x > y \land y' = x) \lor (x < y \land \eq{y}))} \label{eq:ex1-neq} \tag{\protect{\ensuremath{\tau_{\ell_2}^{\neq}}}}
  \end{align}
  At $\ell_1$, $x$ is incremented until $x$ reaches $z$.
  Then, $x$ and $y$ are incremented until $y$ reaches $2 \cdot z + 1$.
  If $x = y = c$ holds for some $c > 1$ at that point, then the execution can continue at $\ell_2$ as follows:
  $
    \ell_2(c,c,c_z) \longrightarrow_{\ref{eq:ex1-eq}} \ell_2(c,c-1,c_z) \longrightarrow_{\ref{eq:ex1-neq}} \ell_2(c-1,c,c_z) \longrightarrow_{\ref{eq:ex1-neq}} \ell_2(c,c,c_z) \longrightarrow_{\ref{eq:ex1-eq}} \ldots
  $
  Here, $\ell_2(c,c,c_z)$ means that the current location is $\ell_2$ and the values of
  $x,y$, and $z$ are $c,c$, and $c_z$.
  (Of course, the value of $z$ could also change arbitrarily in the transitions
  $\ref{eq:ex1-eq}$ and $\ref{eq:ex1-neq}$.)
  Thus, $\TT$ does not terminate.
\end{example}

\Cref{ex:leading} is challenging for state-of-the-art tools for several reasons.
First, more than 5000
steps are required to reach $\ell_2$.
Thus, {\tt chc-LIA-Lin\_052} is beyond the capabilities of most other
state-of-the-art tools for proving reachability.
Second, the pattern ``$\ref{eq:ex1-eq}$, $1^{st}$ disjunct of $\ref{eq:ex1-neq}$, $2^{nd}$ disjunct of $\ref{eq:ex1-neq}$'' must be found to prove non-termination.
Therefore, {\tt flip2\_rec.jar-obl-8} cannot be solved by
other state-of-the-art termination tools.

We present an approach that can prove non-termination of systems
like \Cref{ex:leading} automatically.
To this end, we tightly integrate non-termination techniques into our
recent \emph{Acceleration Driven Clause Learning (ADCL)}
calculus \cite{sas23}, which has originally been designed for Constrained Horn Clauses (CHCs), but it can also be used to analyze TSs.

%% file: preliminaries.tex
\section{Preliminaries}
\label{sec:preliminaries}

We assume familiarity with basics from many-sorted first-order logic.
$\VV$ is a countably infinite set of variables and $\AA$ is a first-order theory over a $k$-sorted signature $\Sigma_\AA$ with carrier $\CC_\AA = (\CC_{\AA,1},\ldots,\CC_{\AA,k})$.
$\QF(\Sigma_\AA)$ is the set of all
quantifier-free first-order formulas over $\Sigma_\AA$, which are  w.l.o.g.\ assumed to be in negation normal form, and $\QF_\land(\Sigma_\AA)$ only contains conjunctions of $\Sigma_\AA$-literals.
Given a first-order formula $\eta$ over $\Sigma_\AA$, $\sigma$ is a \emph{model} of
$\eta$ (written $\sigma \models_\AA \eta$) if it is a model of $\AA$ with carrier
$\CC_\AA$, extended with interpretations for $\VV$ such that $\eta$ is satisfied.
As usual, $\eta \equiv_\AA \eta'$ means $\models_\AA \eta \iff \eta'$.
We write $\vec{x}$ for sequences and $x_i$ is the $i^{th}$ element of $\vec{x}$.
We use ``$::$'' for concatenation of sequences, where we identify sequences of length $1$ with their elements, so we may write, e.g., $x::\mathit{xs}$ instead of $[x]::\mathit{xs}$.
\smallskip

\noindent
{\bf Transition Systems: }
Let $d \in \NN$ be fixed, and let $\vec{x},\vec{x}' \in \VV^d$ be disjoint vectors of pairwise different variables.
Each $\psi \in \QF(\Sigma_\AA)$ induces a relation $\longrightarrow_\psi$ on $\CC_\AA^d$ where $\vec{s} \longrightarrow_\psi \vec{t}$ iff $\psi[\vec{x}/\vec{s},\vec{x}'/\vec{t}]$ is satisfiable.
So for the condition $\psi \Def ({x = y} \land {x > 0} \land \eq{x} \land \dec{y})$
 of \ref{eq:ex1-eq}, we have $(4,4,4)  \longrightarrow_{\psi} (4,3,7)$.
$\LL \supseteq \{\init,\err\}$ is a finite set of \emph{locations}.
A \emph{configuration} is a pair $(\ell,\vec{s}) \in \LL \times \CC_\AA^d$, written $\ell(\vec{s})$.
A \emph{transition} is a triple $\tau = (\ell,\psi,\ell') \in \LL \times \QF(\Sigma_\AA)
\times \LL$, written $\ell \to \ell' \guard{\psi}$, and its \emph{condition} is
$\cond(\tau) \Def \psi$.
W.l.o.g., we assume $\ell \neq \err$ and $\ell' \neq \init$.
Then $\tau$ induces a relation $\longrightarrow_{\tau}$ on configurations where $\state{s} \longrightarrow_{\tau} \state{t}$ iff $\state{s} = \ell(\vec{s}), \state{t} = \ell'(\vec{t})$, and $\vec{s} \longrightarrow_\psi \vec{t}$.
So, e.g., $\ell_2(4,4,4) \longrightarrow_{\tau_{\ell_2}^=} \ell_2(4,3,7)$.
We call $\tau$ \emph{recursive} if $\ell = \ell'$, \emph{conjunctive} if $\psi \in \QF_\land(\Sigma_\AA)$, \emph{initial} if $\ell = \init$, and \emph{safe} if $\ell' \neq \err$.
Moreover, we define $(\ell \to \ell' \guard{\psi})|_{\psi'} \Def \ell \to \ell' \guard{\psi'}$.
A \emph{transition system} (TS) $\TT$ is a finite set of transitions, and it induces the relation $\longrightarrow_{\TT} \Def \bigcup_{\tau \in \TT} {\longrightarrow_{\tau}}$.

\emph{Chaining} $\tau = \ell_s \to \ell_t \guard{\psi}$ and $\tau' = \ell_s' \to \ell_t' \guard{\psi'}$ yields $\chain(\tau,\tau') \Def (\ell_s \to \ell'_t \guard{\psi_c})$ where $\psi_c \Def \psi[\vec{x}' / \vec{x}''] \land \psi'[\vec{x} / \vec{x}'']$ for fresh $\vec{x}'' \in \VV^d$ if $\ell_t = \ell_s'$, and $\psi_c \Def \bot$ (meaning $\mathit{false}$) if $\ell_t \neq \ell_s'$.
So ${\longrightarrow_{\chain(\tau,\tau')}} = {\longrightarrow_\tau} \circ {\longrightarrow_{\tau'}}$, and $\chain(\ref{eq:ex1-nonrec},\ref{eq:ex1-eq}) = \ell_1
\to \ell_2 \guard{\psi}$ where $\psi \equiv_\AA (x = y \land x > 2
\cdot z \land x > 0 \land \eq{x} \land \dec{y})$.
For non-empty, finite sequences of transitions we define $\chain([\tau]) \Def \tau$ and
$\chain([\tau_1,\tau_2]::\vec{\tau}) \Def \chain(\chain(\tau_1,\tau_2)::\vec{\tau})$.
We lift notations for transitions to finite sequences via chaining.
So $\cond(\vec{\tau}) \Def \cond(\chain(\vec{\tau}))$,
$\vec{\tau}$ is \emph{recursive} if $\chain(\vec{\tau})$ is recursive,
${\longrightarrow_{\vec{\tau}}} = {\longrightarrow_{\chain(\vec{\tau})}}$, etc.
If $\tau$ is initial and $\cond(\tau::\vec{\tau}) \not\equiv_\AA \bot$, then $(\tau::\vec{\tau}) \in \TT^+$ is a \emph{finite run}.
$\TT$ is safe if every finite run is safe.
If there is a $\sigma$ such that $\sigma \models_\AA \cond(\vec{\tau}')$ for every finite prefix $\vec{\tau}'$ of $\vec{\tau} \in \TT^\omega$, then $\vec{\tau}$ is an \emph{infinite run}.
If no infinite run exists, then $\TT$ is \emph{terminating}.
\medskip

\noindent
{\bf Acceleration Techniques: }
\emph{Acceleration techniques} compute transitive closures of relations.

\begin{definition}[Acceleration]
  \label{def:accel}
  An \emph{acceleration technique} is a function $\accel: \QF_\land(\Sigma_\AA) \mapsto
  \QF_\land(\Sigma_{\AA'})$ such that ${\longrightarrow_{\psi}^+} =
     {\longrightarrow_{\accel(\psi)}}$,
     where $\AA'$ is a first-order theory.
  For recursive conjunctive transitions $\tau$, we define $\accel(\tau) \Def \tau|_{\accel(\cond(\tau))}$.
\end{definition}
\Cref{def:accel} allows $\AA' \neq \AA$ as most theories are not ``closed under acceleration''.
E.g., accelerating the linear formula $x'_1 = x_1 + x_2 \land \eq{x_2}$ yields $n > 0 \land x'_1 = x_1 + n \cdot x_2 \land \eq{x_2}$, which is non-linear.

%% file: calculus.tex
\section{Proving Non-Termination with ADCL}
\label{sec:ADCL}

To bridge the gap between transitions $\tau$ where $\cond(\tau) \in \QF(\Sigma_\AA)$ and acceleration techniques for formulas from $\QF_\land(\Sigma_\AA)$, ADCL uses \emph{syntactic implicants}.

\begin{definition}[Syntactic Implicants \protect{\cite[Def.\ 6]{sas23}}]
  \label{def:implicant}
  If $\psi \in \QF(\Sigma_\AA)$, then:
  \begin{align*}
    \mbp(\psi,\sigma) & {} \Def \bigwedge \{\pi \text{ is a literal of } \psi \mid \sigma \models_\AA \pi\} && \text{if $\sigma \models_\AA \psi$} \\
    \mbp(\psi) & {} \Def \{ \mbp(\psi,\sigma) \mid \sigma \models_\AA \psi \}\\
    \mbp(\tau) & {} \Def \{\tau|_{\psi} \mid \psi \in \mbp(\cond(\tau))\} && \text{for transitions $\tau$} \\
    \mbp(\TT) & {} \Def \bigcup_{\tau \in \TT} \mbp(\tau) && \text{for TSs $\TT$}
  \end{align*}
  Here, $\mbp$ abbreviates \emph{syntactic implicant projection}.
\end{definition}
While $\mbp(\psi)$ contains  $\mbp(\psi,\sigma)$ for all models
$\sigma$ of $\psi$, the set $\mbp(\psi)$ is finite, because
$\mbp(\psi,\sigma)$ is restricted to literals from $\psi$.
Syntactic implicants ignore the semantics of literals.
So we have, e.g., $(X > 1) \notin \mbp(X > 0 \land X > 1) = \{X > 0 \land X > 1\}$.
It is easy to show $\psi \equiv_\AA \bigvee \mbp(\psi)$, and thus ${\longrightarrow_\TT} = {\longrightarrow_{\mbp(\TT)}}$.

The core idea of ADCL is to learn new, \emph{non-redundant} transitions via acceleration.

\begin{definition}[Redundancy, \protect{\cite[Def.\ 8]{sas23}}]
  \label{def:redundancy}
  A transition $\tau$ is \emph{(strictly) redundant} w.r.t.\ $\tau'$, denoted $\tau \sqsubseteq \tau'$ ($\tau \sqsubset \tau'$) if ${\longrightarrow_\tau} \subseteq {\longrightarrow_{\tau'}}$ (${\longrightarrow_\tau} \subset {\longrightarrow_{\tau'}}$).
  For a TS $\TT$, we have $\tau \sqsubseteq \TT$ ($\tau \sqsubset \TT$) if $\tau \sqsubseteq \tau'$ ($\tau \sqsubset \tau'$) for some $\tau' \in \TT$.
\end{definition}
To prove non-termination, we look for a corresponding \emph{certificate}.
\begin{definition}[Certificate of Non-Termination]
  \label{def:certificate}
  Let $\tau = \ell \to \ell \guard{\ldots}$.
  A satisfiable formula $\psi$ \emph{certifies non-ter\-mi\-na\-tion of $\tau$}, written
  $\psi \models_\AA^\infty \tau$, if for any model $\sigma$ of $\psi$,
  there is an infinite sequence
  $
    \ell(\sigma(\vec{x})) = \state{s}_1 \longrightarrow_\tau \state{s}_2 \longrightarrow_\tau \ldots
  $
\end{definition}
From now on, let $\TT$ be the TS that is being analyzed with ADCL, and assume that $\TT$ does not contain unsafe transitions.
A \emph{state} of ADCL consists of a TS $\SS$ that augments $\TT$ with
\emph{learned transitions}, a run $\vec{\tau}$ of $\SS$ called the
\emph{trace},
and a sequence of sets of \emph{blocking transitions} $[B_i]_{i=0}^k$,
where transitions that are redundant w.r.t.\ $B_k$ must not be appended to the trace.

\begin{definition}[ADCL]
  \label{def:state}
  \label{def:calc}
  A \emph{state} is a triple $(\SS,[\tau_i]_{i=1}^k,[B_i]_{i=0}^k)$
  where $\SS \supseteq \TT$ is a TS, $\bigcup_{i=0}^k B_i \subseteq \mbp(\SS)$, and $[\tau_i]_{i=1}^k \in \mbp(\SS)^*$.
  The transitions in $\mbp(\TT)$ are called \emph{original} and the transitions in $\mbp(\SS)\setminus\mbp(\TT)$ are \emph{learned}.
  A transition $\tau_{k+1} \sqsubseteq B_k$ is \emph{blocked}, and 
  $\tau_{k+1} \not\sqsubseteq B_k$ is \emph{active} if $\chain([\tau_i]_{i=1}^{k+1})$ is
  an initial transition with satisfiable condition (i.e., $[\tau_i]_{i=1}^{k+1}$ is a run). Let
  $
    \bt(\SS,[\tau_i]_{i=1}^k,[B_0,\ldots,B_k]) \Def (\SS,[\tau_i]_{i=1}^{k-1},[B_0,\ldots,B_{k-1} \cup \{\tau_k\}]),
  $
  where $\bt$ abbreviates ``backtrack''.
  Our calculus is defined by the following rules.

  \medskip\noindent
  \scalebox{0.9}{
  \[
    \begin{array}{cr@{\quad}cr}
      \infer{\TT \leadsto (\TT,[],[\emptyset])}{} & (\Init) & \infer{(\SS,\vec{\tau},\vec{B}) \leadsto (\SS,\vec{\tau}::\tau,\vec{B}::\emptyset)}{\tau \in \mbp(\SS) \text{ is active}} & (\Step)
      \\[0.8em]
      \multicolumn{3}{c}{\infer{(\SS,\vec{\tau}::\vec{\tau}^\circlearrowleft,\vec{B}::\vec{B}^\circlearrowleft) \leadsto (\SS \cup \{\tau\},\vec{\tau}::\tau,\vec{B}::\{\tau\})}{\vec{\tau}^\circlearrowleft \text{ is recursive} & |\vec{\tau}^\circlearrowleft| = |\vec{B}^\circlearrowleft| & \accel(\vec{\tau}^\circlearrowleft) = \tau \not\sqsubseteq \mbp(\SS)}} & \mathllap{(\Accelerate)} \\[0.8em]
      \multicolumn{3}{c}{\infer{
      s = (\SS,\vec{\tau}::\vec{\tau}^\circlearrowleft,\vec{B}) \leadsto \bt(s)
      }{\vec{\tau}^\circlearrowleft \text{ is recursive} & \quad \vec{\tau}^\circlearrowleft \sqsubset \mbp(\SS) \text{ or } \vec{\tau}^\circlearrowleft \sqsubseteq \mbp(\SS) \land |\vec{\tau}^\circlearrowleft| > 1}} & (\Covered) \\[0.8em]
      \infer{(\SS,\vec{\tau},\vec{B}) \leadsto \unsafe}{\vec{\tau} \text{ is unsafe}} & (\Refute) & \infer{s = (\SS,\vec{\tau}::\tau,\vec{B}) \leadsto \bt(s)}{\text{all transitions from $\mbp(\SS)$ are inactive} & \tau \text{ is safe}} & \mathllap{(\Backtrack)} \\[0.8em]
      \multicolumn{3}{c}{\infer{
      (\SS,\vec{\tau}::\vec{\tau}^\circlearrowleft,\vec{B}) \leadsto (\SS \cup \{\tau\}, \vec{\tau}::\vec{\tau}^\circlearrowleft,\vec{B})
      }{\chain(\vec{\tau}^\circlearrowleft) = \ell \to \ell \guard{\ldots} & \psi \models_\AA^\infty \vec{\tau}^\circlearrowleft & \tau = \ell \to \err \guard{\psi} \not\sqsubseteq \mbp(\SS)}} & (\Nonterm)
    \end{array}
  \]
  }
\end{definition}
We write $\overset{\textsc{I}}{\leadsto}$, $\overset{\textsc{S}}{\leadsto}$, $\ldots$ to indicate that the rule \Init, {\Step}, $\ldots$ was used.
{\Step} adds a transition to the trace.
When the trace has a recursive suffix, \Accelerate allows for learning a new transition which replaces the recursive suffix on the trace, or we may backtrack via \Covered if the recursive suffix is redundant.
Note that \Covered does not apply if $\vec{\tau}' \sqsubseteq \mbp(\SS)$ and $|\vec{\tau}'| = 1$, as it could immediately undo every \Step, otherwise.
If no further {\Step} is possible, \Backtrack applies.
Note that \Backtrack and \Covered block the last transition from the trace so that we do not perform the same {\Step} again.
If $\vec{\tau}$ is unsafe, \Refute yields $\unsafe$.
As $\TT$ is safe, this only happens if \Nonterm, which applies a non-termination technique to a recursive suffix of the trace, added an unsafe transition before.

\begin{example}
  \label{ex:unsafe}
  We apply ADCL to \Cref{ex:leading}
      \begin{align*}
        \TT \overset{\textsc{I}}{\leadsto}^{\phantom{2}} {} & (\TT,[],[\emptyset]) \overset{\textsc{S}}{\leadsto}^2 (\TT,[\ref{eq:ex1-init},\ref{eq:ex1-rec}|_{\psi_{x < z}}],[\emptyset,\emptyset,\emptyset]) \tag{$x \leq 1 \land z \geq 5k \land y \leq z$} \\
        {} \overset{\textsc{A}}{\leadsto}^{\phantom{2}} {} & (\SS_1,[\ref{eq:ex1-init},\tau_{x<z}^+],[\emptyset,\emptyset,\{\tau_{x<z}^+\}]) \tag{$x \leq z \land z \geq 5k \land y \leq z$} \\
        {} \overset{\textsc{S}}{\leadsto}^{\phantom{2}} {} & (\SS_1,[\ref{eq:ex1-init},\tau_{x<z}^+,\ref{eq:ex1-rec}|_{\psi_{x \geq z}}],[\emptyset,\emptyset,\{\tau_{x<z}^+\},\emptyset]) \tag{$x = z + 1 \land z \geq 5k \land y \leq z + 1$} \\
        {} \overset{\textsc{A}}{\leadsto}^{\phantom{2}} {} & (\SS_2,[\ref{eq:ex1-init},\tau_{x<z}^+,\tau_{x \geq z}^+],[\emptyset,\emptyset,\{\tau_{x<z}^+\},\{\tau_{x \geq z}^+\}]) \tag{$x \geq y \land x > z \geq 5k \land y \leq 2 \cdot z + 1$} \\
        {} \overset{\textsc{S}}{\leadsto}^4_\nt
        {} & (\SS_2,[\ref{eq:ex1-init},\tau_{x<z}^+,\tau_{x \geq
            z}^+,\ref{eq:ex1-nonrec},\ref{eq:ex1-eq},\ref{eq:ex1-neq}|_{\psi_{x>y}},\ref{eq:ex1-neq}|_{\psi_{x<y}}],[\ldots])
        \tag{$1 \equiv_2 y = x > 10k \land \ldots$} \\
        {} \overset{\textsc{N}}{\leadsto}_\nt {} &
        (\SS_3,[\ref{eq:ex1-init},\tau_{x<z}^+,\tau_{x \geq
            z}^+,\ref{eq:ex1-nonrec},\ref{eq:ex1-eq},\ref{eq:ex1-neq}|_{\psi_{x>y}},\ref{eq:ex1-neq}|_{\psi_{x<y}}],[\ldots])
        \tag{$1 \equiv_2 y = x > 10k \land \ldots$} \\
        {} \overset{\textsc{S}}{\leadsto}_\nt {} & (\SS_3,[\ref{eq:ex1-init},\tau_{x<z}^+,\tau_{x \geq z}^+,\ref{eq:ex1-nonrec},\ref{eq:ex1-eq},\ref{eq:ex1-neq}|_{\psi_{x>y}},\ref{eq:ex1-neq}|_{\psi_{x<y}},\tau_\err],[\ldots]) \overset{\textsc{R}}{\leadsto}_\nt {} \unsafe
      \end{align*}
  Here, $5k$ abbreviates $5000$ and:

  \medskip\noindent
  \scalebox{0.95}{
  \[
  \begin{aligned}
    \psi_{x < z} & {} \Def y \leq 2 \cdot z \land \inc{x} \land x < z \land \eq{y} \land \eq{z} & \psi_{x \geq z} & {} \Def y \leq 2 \cdot z \land \inc{x} \land x \geq z \land \inc{y} \land \eq{z} \\
    \tau_{x<z}^+ & {} \Def \mathrlap{\ell_1 \to \ell_1 \guard{y \leq 2 \cdot z \land n > 0 \land x' = x + n \land x + n \leq z \land \eq{y} \land \eq{z}}} \\
    \tau_{x \geq z}^+ & {} \Def \mathrlap{\ell_1 \to \ell_1 \guard{y + n - 1 \leq 2 \cdot z \land n > 0 \land x' = x + n \land x \geq z \land y' = y + n \land \eq{z}}} \\
    \psi_{x>y} & {} \Def x > 0 \land y > 0 \land x' = y \land x > y \land y' = x & \psi_{x<y} & {} \Def x > 0 \land y > 0 \land x' = y \land x < y \land \eq{y} \\
    \SS_1 & {} \Def \TT \cup \{\tau_{x<z}^+\} \hspace{2em} \mathrlap{\SS_2 \Def \SS_1 \cup \{\tau_{x \geq z}^+\}  \hspace{2em} \SS_3 \Def \SS_2 \cup \{\tau_\err\} \hspace{2em}
    \tau_\err \Def \ell_2 \to \err \guard{x = y > 1}}
  \end{aligned}
  \]
  }

  \medskip\noindent
  On the right, we show formulas describing the configurations that are reachable with the
  current trace, where $1 \equiv_2 y$ means that $y$ is odd. 
  Every $\leadsto$-derivation starts with \Init.
  The first two {\Step}s add the initial transition \ref{eq:ex1-init} and an element of $\mbp(\ref{eq:ex1-rec})$ to the trace.
  Since $x < z$ holds after applying \ref{eq:ex1-init}, the only possible choice for the latter is $\ref{eq:ex1-rec}|_{\psi_{x < z}}$.

  As $\ref{eq:ex1-rec}|_{\psi_{x < z}}$ is recursive, it is accelerated and replaced with $\accel(\ref{eq:ex1-rec}|_{\psi_{x < z}}) = \tau^+_{x<z}$, which simulates $n$ steps with $\ref{eq:ex1-rec}|_{\psi_{x < z}}$.
  Moreover, $\tau_{x<z}^+$ is also added to the current set of blocking transitions, as we
  always have ${\longrightarrow^2_{\tau}} \subseteq {\longrightarrow_{\tau}}$ for learned
  transitions $\tau$ and thus adding them to the trace twice in a row is pointless.

  Next, \ref{eq:ex1-rec} is applicable again.
  As neither $x < z$ nor $x \geq z$ holds for all reachable configurations, we could continue with any element of $\mbp(\ref{eq:ex1-rec}) = \{\ref{eq:ex1-rec}|_{\psi_{x < z}},\ref{eq:ex1-rec}|_{\psi_{x \geq z}}\}$.
  We choose $\ref{eq:ex1-rec}|_{\psi_{x \geq z}}$, so that the recursive transition $\ref{eq:ex1-rec}|_{\psi_{x \geq z}}$ can be accelerated to $\tau^+_{x \geq z}$.

  After the next \Step with \ref{eq:ex1-nonrec}, just \ref{eq:ex1-eq} can be used, as $\cond(\ref{eq:ex1-nonrec})$ implies $x' = y'$.
  While \ref{eq:ex1-eq} is recursive, \Accelerate cannot be applied next, as ${{\longrightarrow_{\ref{eq:ex1-eq}}} = {\longrightarrow_{\ref{eq:ex1-eq}}^+}}$, so the learned transition would be redundant.
  Thus, we continue with \ref{eq:ex1-neq}, projected to $x>y$ (as $\cond(\ref{eq:ex1-eq})$ implies $x' = y'+1$).
  Again, all transitions that could be learned are redundant, so \Accelerate does not apply.
  We next use \ref{eq:ex1-neq} projected to $x<y$, as the previous \Step swapped $x$ and $y$.
  As the suffix
  $[\ref{eq:ex1-eq},\ref{eq:ex1-neq}|_{\psi_{x>y}},\ref{eq:ex1-neq}|_{\psi_{x<y}}]$ of the
  trace does not terminate (see \Cref{ex:leading}), \Nonterm applies.
  So we learn the transition $\tau_\err$, which is added to the trace to finish the proof, afterwards.  
\end{example}
\begin{restatable}{theorem}{correctnt}
  \label{thm:sound-nt}
  If $\TT \leadsto_\nt^* \unsafe$, then $\TT$ does not terminate.
\end{restatable}
See \cite{report} for a discussion of obstacles regarding an adaption of ADCL for \emph{proving} termination.

%% file: experiments.tex
\section{Implementation and Experiments}
\label{sec:experiments}

So far, our implementation in our tool \tool{LoAT} is restricted to integer arithmetic.
It uses the technique from \cite{loat} for acceleration and finding certificates of non-termination, the SMT solvers \tool{Z3} \cite{z3} and \tool{Yices} \cite{yices}, the recurrence solver \tool{PURRS} \cite{purrs}, and \tool{libFAUDES} (\url{https://fgdes.tf.fau.de/faudes}) to implement the automata-based redundancy check from \cite{sas23}.

To evaluate our implementation in \tool{LoAT}, we used the 1222 \emph{Integer Transition Systems} (ITSs) from the \emph{Termination Problems
Database} (\url{https://termination-portal.org/wiki/TPDB}) used in \emph{TermComp} \cite{termcomp}.
We compared our implementation (\tool{LoAT ADCL}) with other leading termina\-tion analyzers: \tool{iRankFinder}~\cite{irankWST}, \tool{T2}~\cite{t2-tool}, \tool{VeryMax}~\cite{larraz14}, and the previous version of \tool{LoAT}~\cite{loat} (\tool{LoAT '22}).
For \tool{T2} and \tool{VeryMax}, we took the versions of their last \emph{TermComp} participations (2015 and 2019).
For \tool{iRankFinder}, we used the configuration from the evaluation of \cite{loat}, which is tailored towards proving non-termination.
All tests were run on \tool{StarExec} with $300$s wallclock timeout, $1200$s CPU timeout,
and $128$GB memory limit per example.

\begin{table}
  \begin{center}
    \begin{tabular}{|c||c|c||c||c|c|c||c|c|}
      \hhline{~--------} \multicolumn{1}{c|}{} & \multicolumn{2}{c||}{No} & Yes & \multicolumn{3}{c||}{Runtime overall} & \multicolumn{2}{c|}{Runtime No} \\
      \hhline{~--------} \multicolumn{1}{c|}{}    & solved & unique & solved & average & median & timeouts & average & median \\
      \hhline{-========} \tool{LoAT ADCL}
                                               & 521    & 9      & 0      & 48.6~s  & 0.1~s  & 183      & 2.9~s   & 0.1~s  \\
      \hhline{---------} \tool{LoAT '22}     & 494    & 2      & 0      & 7.4~s   & 0.1~s  &   0      & 6.2~s   & 0.1~s  \\
      \hhline{---------} \tool{T2}           & 442    & 3      & 615    & 17.2~s  & 0.6~s  &  45      & 7.4~s   & 0.6~s  \\
      \hhline{---------} \tool{VeryMax}      & 421    & 6      & 631    & 28.3~s  & 0.5~s  &  30      & 30.5~s  & 14.5~s \\
      \hhline{---------} \tool{iRankFinder}  & 409    & 0      & 642    & 32.0~s  & 2.0~s  &  93      & 12.3~s  & 1.7~s  \\
      \hline
    \end{tabular}
  \end{center}
\end{table}

The table above shows the results of our experiments, where the column ``unique'' contains the number of examples that could be solved by
the respective tool, but no others.
It shows that \tool{LoAT ADCL} is the most powerful tool for proving non-termination of ITSs.

If we only consider the examples where non-termination is proven, \tool{LoAT ADCL} is also the fastest tool.
If we consider all examples, then the \emph{average} runtime of \tool{LoAT ADCL} is significantly slower.
This is not surprising, as ADCL does not terminate in general \cite[Thm.~18]{sas23}.
So while it is very fast in most cases (as witnessed by the very fast \emph{median} runtime), it times out more often than the other tools.
Note that \tool{LoAT ADCL} does not subsume \tool{LoAT '22}.
The reason is that \tool{LoAT '22} under-approximates more aggressively and hence solves some instances where \tool{LoAT ADCL} times out.

See \cite{website} for detailed results and a pre-compiled binary.
\tool{LoAT} is open-source and available on GitHub: \url{https://github.com/LoAT-developers/LoAT}